\documentclass[aps,english,prd,showpacs,showkeys,twocolumn,longbibliography, nofootinbib]{revtex4-1}

\usepackage[english]{babel}
\usepackage[utf8]{inputenc}
\usepackage[T1]{fontenc}
\usepackage{fnpct} % for multiple footnotes
\usepackage{ulem} % for \bar texts
\usepackage{scrextend} % for multiple footnote ref
\usepackage{enumitem}

\usepackage{amsmath}
\usepackage{amsfonts}
\usepackage{amssymb}
\usepackage{mathtools, stmaryrd}
\usepackage{empheq}
\usepackage{tensor}

\usepackage{fancyvrb}
\usepackage{graphicx}
\usepackage{floatrow}
\usepackage[usenames,dvipsnames]{xcolor}
\usepackage{tabularx}
	\newcolumntype{x}[1]{>{\centering\arraybackslash\hspace{0pt}}p{#1}} % For centered column with specific width
\usepackage{caption}
	\DeclareCaptionJustification{justified}{\leftskip=0pt \rightskip=0pt \parfillskip=0pt plus 1fil}
\usepackage{float}
\floatstyle{plaintop}
\restylefloat{table}

\usepackage{url}
\usepackage[pdftex,colorlinks=true, pdfstartview=FitV, linkcolor= Mlink, citecolor=Mlink, urlcolor= Mlink, hyperindex=true, hyperfigures=true]{hyperref}
\hypersetup{linktoc=page}

\IfFileExists{aastexv6.sty}{\input{aastexv6.sty}}{\input{/Users/quentinvigneron/Documents/Travail/Research/tex_/aastexv6.sty}}% For ADS abbreviations
\IfFileExists{Short_cut.sty}{\input{Short_cut.sty}}{\input{/Users/quentinvigneron/Documents/Travail/Research/tex_/Short_cut.sty}}

\begin{document}

\title[Gravitational potential in spherical topologies]{Gravitational potential in spherical topologies}

\author{Quentin Vigneron$^{1}$, Boudewijn Roukema$^{1,2}$}
\address{$^1$ Institute of Astronomy, Faculty of Physics, Astronomy and Informatics, Nicolaus Copernicus University, Grudziadzka 5, 87-100 Toru\'n, Poland}
\address{$^2$ Univ Lyon, Ens de Lyon, Univ Lyon1, CNRS, Centre de Recherche Astrophysique de Lyon UMR5574, F--69007, Lyon, France}
\email{quentin.vigneron@umk.pl}
\email{boud astro.uni.torun.pl}

\vspace{10pt}
\date{\today}

\begin{abstract}

We study the properties of the Newtonian gravitational potential in a spherical Universe for different topologies. For this, we use the non-Euclidean Newtonian theory developed in Vigneron [\href{https://ui.adsabs.harvard.edu/abs/2022CQGra..39o5006V/abstract}{2022, \cqg, {\bf 39}, 155006}] describing Newtonian gravitation in a spherical or hyperbolic Universe.
The potential is calculated for a point mass in all the globally homogeneous regular spherical topologies, i.e. whose fundamental domain is unique and is a platonic solid.
We provide the exact solution and the Taylor expansion series of the potential at a test position near the point mass. We show that the odd terms of the expansion can be interpreted as coming from the presence of a non-zero spatial scalar curvature, while the even terms relate to the closed nature of the topological space. A consequence is that, compared to the point mass solution in a 3-torus, widely used in Newtonian cosmological simulations, the spherical cases all feature an additional attractive first order term dependent solely on the spatial curvature. The choice of topology only affects the potential at second order and higher. For typical estimates of cosmological scales (curvature and topology), the strongest topological effect occurs in the case of the Poincar\'e dodecahedral space, but in general the effect of curvature dominates over topology. We also provide the set of equations that can be used to perform $N$-body simulations of structure formation in spherical topologies.

\end{abstract}

\maketitle

\section{Introduction}

In the $\Lambda$CDM model, three cosmological expansion scenarios are possible depending on the Thurston topological class adopted for the spatial 3-manifold: spherical, Euclidean, or hyperbolic \cite{1982_Thurston} (the other five classes, which forbid local isotropy, are not described by the $\Lambda$CDM model).
Each of these classes corresponds to an ensemble of topological spaces whose covering space is, respectively, the 3-sphere $\mS^3$, the Euclidean space $\mE^3$, or the hyperbolic space $\mH^3$.
However, due to the homogeneity hypothesis of the model, the specific choice of topology within a class does not affect the global expansion. In contrast, taking into account the presence of inhomogeneities allows for the search of the specific topology (e.g. multiply connected) of our Universe by searching for correlations of matter distributions using either catalogues of extragalactic objects (the methods of cosmic crystallography, e.g., \citep[][]{1996_Roukema, 1996_Le_La_Lu, 2011_Fujii_et_al_a}), or the cosmic microwave background (CMB) map (the method of circles in the sky, \cite[][]{1998_Cornish_et_al}). These studies currently give typical lower bounds of around (11--18 Gpc$/h$)$^3$ (e.g. \cite{2014_Roukema_et_al, 2014_Planck_XXVI, 2015_Planck_XVIII, 2020_Planck_VI}, where $h \coloneqq H_0/100$~km/s/Mpc is the dimensionless Hubble--Lema\^{\i}tre constant) for the comoving volume of our Universe for the 3-manifolds studied so far (hereafter, ``topologies''; see Sec.~\ref{sec:top_terminology}) . These methods are all based on some form of the spatial correlations of the matter distribution projected to a single (early or late) time slice. Thus, they do not probe the potential effects of topology on the dynamics, either global \citep{2020_Brunswic_et_al, 2022_Vigneron} or local \citep{1990_Farrar_et_al, 2007_Roukema_et_al}, of our Universe.

The effects of topology and curvature on structure formation are generally expected to be weak in our Universe. The reason is that the homogeneity scale ($\sim100$ Mpc/h, e.g. \citep[][]{2018_Goncalves_et_al}) is much smaller than the recent estimates of the lower bounds (of a few gigaparsecs) for the finite size of our Universe and its curvature radius. However, the constraints on topology are not general but are limited to a small subset of possible topologies \citep{2022_Akrami_et_al}; and an increasing debate on the value of the spatial curvature preferred from the data has arisen during the past few years (e.g. \citep[][]{2020_Di-Valentino_et_al, 2020_Efstathiou_et_al, 2021_Di-Valentino_et_al_bis, 2021_Handley}). Therefore in a era of precision cosmology, performing non-linear structure formation simulations in different (non-Euclidean) topologies (other than the 3-torus currently considered in most cosmological simulations) remains of interest to properly quantify the role, especially in the shape parameters of the structures, that can be attributed to topology and/or curvature.

The non-linear structures in our Universe being non-relativistic, i.e. having velocity dispersions small compared to the speed of light, a non-relativistic calculation/simulation is generally considered precise enough to describe the dynamics of these structures. This is one of the reasons why we currently use Newtonian cosmological simulations to study them. However, Newton's theory being defined on a Euclidean topology (i.e. on a 3-manifold whose covering space is $\mE^3$, taken to be the 3-torus in simulations), that theory cannot be used to study structure formation in non-Euclidean (e.g. spherical) topologies. Therefore, until recently, the only theory that could allow for such a study was general relativity. However, simulations that directly solve the Einstein equation in a cosmological context (e.g. \cite{2019_Macpherson_et_al} and references therein) are still far from reaching the precision that is possible with Newtonian $N$-body simulations and needed to fully take into account the non-linear regime.

In \cite{2022_Vigneron_b}, we extended the validity of Newton's theory of gravity to spherical and hyperbolic topologies with the aim of describing the non-relativistic regime in these spaces. That theory, called \textit{non-Euclidean Newtonian theory} (NEN theory), is to be understood as \textit{an extension} (to include non-Euclidean topologies) and \textit{not a modification} of Newtonian gravitation, since Newton's second law is still valid, contrary to MOND theory. As for Newton's theory, the NEN theory allows for the encoding of non-linearities and all the global properties (i.e. topology) of general relativity, while being simpler to use. In particular, an exact $N$-body description of gravity for point masses exists in this theory, making ``fast'' (non-relativistic) $N$-body cosmological simulations in spherical or hyperbolic topologies practical. Performing such simulations requires knowledge of the point-mass gravitational potential in these topologies.

The goal of this paper is to calculate this potential and analyse its properties by solving the NEN equations in different topologies, focusing for now on the spherical ones, and thus paving the way for future structure formation simulations in these topologies. This extends beyond the work of \citep{2007_Roukema_et_al}, as our NEN theory is better justified and does not feature physical inconsistencies (see Sec.~\ref{sec:NEN_phil}). Our approach also differs from \citep{2019_Eingorn_et_al}, who calculated the gravitational potential in a spherical universe, the main differences being that we do not introduce a screening length and we consider multiconnected spherical topologies, i.e. not only the 3-sphere.

In Sec.~\ref{sec:NEN}, we summarise the system of equations of the NEN theory \cite{2022_Vigneron_b}, and simplify it for the case of a single point mass in a spherical topology. Section~\ref{sec:pot} presents our characterisation of the ``regular'' spherical topologies and the Taylor expansion series of the gravitational potential near the point mass in each of these topologies. We interpret these results in Sec.~\ref{sec:disc} and conclude in Sec.~\ref{sec:ccl}.

\section{Non-Euclidean Newtonian theory}
\label{sec:NEN}

\subsection{What is NEN theory and how is it constructed?}
\label{sec:NEN_phil}

A NEN theory is a theory which is invariant under local Galilean transformations (hence ``Newtonian'', or equivalently ``non-relativistic'') and defined on a non-Euclidean topology (see Sec.~\ref{sec:top_terminology}, Sec. IV.A in \citep{2022_Vigneron}, and Sec.~2 in \citep{2022_Vigneron_b} for a precise definition of that term). A non-Euclidean topology necessarily has a non-zero spatial Ricci tensor. The purpose of this theory is to describe the non-relativistic regime of a Universe having a non-Euclidean topology (of which the spherical and hyperbolic cases are of most interest), in the same way Newton's theory describes this regime in a Euclidean topology. Therefore, this extension of Newton's theory does not aim at taking into account post-Newtonian terms or effects of spatial curvature that would come from general relativity (as in e.g. \citep{2014_Abramowicz_et_al}), nor does it modify Newton's second law of gravitation as in MOND theory [see Eq.~\eqref{eq::NEN_def_g_2}].

The study of topology within a non-relativistic theory is possible because, fundamentally, such a theory is constructed on a 4-manifold (as in general relativity), which is the mathematical object carrying the topological property \citep{2021_Vigneron, 2022_Vigneron_b}. In other words, the notion of topology does not only belong to Lorentz invariant theories (i.e. relativistic), but also to Galilean invariant theories (i.e. non-relativistic).

A prototype of NEN theory was already proposed before that of Ref.~\cite{2022_Vigneron_b} by Refs.~\cite{2009_Roukema_et_al, 2020_Barrow}, based on the introduction of a non-zero spatial curvature into the Poisson equation. However, that theory suffers from two major problems: it cannot describe cosmological expansion, and, in spherical topologies, the gravitational field of a point mass is necessarily matched by a white hole (a repulsive singular gravitational field) at the antipode of the point mass, making the theory physically dubious (see Sec.~4.2 in \cite{2022_Vigneron_b} for a detailed discussion of the problems). The approach of \cite{2009_Roukema_et_al, 2020_Barrow} is to consider the Poisson equation as a fundamental feature of a non-relativistic theory, regardless of the topology.

In \cite{2022_Vigneron_b}, we instead constructed a NEN theory where Galilean invariance is considered to be a fundamental principle of a non-relativistic theory no matter the topology. This has been made possible by using the concept of Galilean manifolds \cite{1972_Kunzle} and a minimal modification of the Newton--Cartan equations. We were able to define two NEN theories with this approach, but only one (that of Sec.~5.6 in \cite{2022_Vigneron_b}) turned out to be physically reasonable, and we argued that it should be considered as the ``right'' extension of Newton's theory for non-Euclidean topologies. This theory solves, in particular, the two problems quoted above that were present in the proposal of \cite{2009_Roukema_et_al, 2020_Barrow}.  The detailed construction of this theory can be found in \cite{2022_Vigneron_b}, showing, in particular, how the 3-dimensional gravitational system of equations can be obtained from the 4-dimensional modified Newton--Cartan equation. Thus, we adopt this gravitational system, i.e. featuring the gravitational field. We present its general form in Sec.~\ref{sec:NEN_grav_gene} and the specific form used for calculations in the current paper in Sec.~\ref{sec:NEN_grav_simp}.

\subsection{General form of the gravitational system in the NEN theory}
\label{sec:NEN_grav_gene}

The gravitational system of the NEN theory as derived by \cite{2022_Vigneron_b} is defined on a \textit{closed} 3-manifold $\Sigma$ whose topology belongs to the class of spherical or hyperbolic topologies of the Thurston classification \footnote{Cosmology textbooks often use the term ``open'' to refer uniquely to hyperbolic curvature and the associated universe expansion history; and ``closed'' to refer to spherical curvature and the associated expansion history. Here, we do not adopt this confusing terminology, and instead use the language of topological manifolds, since general-relativistic cosmology requires the Universe to be a pseudo-Riemannian 4-manifold.}(i.e. whose covering space is either $\mS^3$ or $\mH^3$). The NEN theory is currently defined only for these two classes, but may be extended in a later study to the remaining five non-Euclidean irreducible classes of 3-dimensional closed topologies of the Thurston classification.

The most general form of the gravitational system in the NEN theory is\footnote{We assumed that the harmonic 2-form $\T\omega$ present in the system~(60)--(66) in \cite{2022_Vigneron_b} is zero. This is expected if these equations result from the non-relativistic limit of general relativity (see Appendix~B in \cite{2021_Vigneron}).}
\begin{align}
        g^a &= \left(\partial_t - \Lie{\T \beta}\right) v^a + v^c\D_c v^a+ 2v^c \left(H {\delta_c}^a + {\Xi_c}^a\right) \nonumber \\&\qquad - (a_{\not= \rm grav})^a, \label{eq::NEN_def_g_1} \\
        \D_c g^c &= -4\pi G \widehat{\rho} - \widehat{\Xi_{cd}\Xi^{cd}} \label{eq::NEN_New_Gaus_1},\\
        \D_{[a} g_{b]} &= 0.\label{eq::NEN_g_2}
\end{align}
These equations are completed by
\begin{align}
        \Rt_{ab} &= \frac{\Rt(t)}{3} h_{ab}\, , \label{eq::NEN_sys_Ri} \\
        \left(\partial_t - \Lie{\T \beta}\right)h_{ab}  &= 2 \left(H h_{ab} + \D_{(a}v_{b)} + \Xi_{ab}\right), \label{eq::NEN_Theta_hb} \\
        \left(\partial_t - \Lie{\T \beta}\right) \rho &= -\rho\left(3H + \D_c v^c\right), \label{eq::NEN_cont}
\end{align}
and the expansion law
\begin{align}
        3\left(\dot H + H^2\right) + 4\pi G \Saverage{\rho} - \Lambda = - \Saverage{\Xi_{cd}\Xi^{cd}}, \label{eq::NEN_sys_av_rho}
\end{align}
with $\Saverage{\rho} = \frac{M_{\rm tot}}{V_{\Sigma}(t)}$ and where
\begin{enumerate}
        \item $\T g$ is the gravitational field.
        \item $\Lie{\T\beta}$ is the Lie derivative with respect to the vector $\T \beta$. This vector is a free parameter that corresponds to a choice of spatial coordinates (see the next section).
        \item $\T v$ is the spatial velocity of the fluid.
        \item $\T D$ is the Levi-Civita connection relative to the metric $\T h$ whose Ricci tensor $\T\CR$ is given by formula~\eqref{eq::NEN_sys_Ri}.
        \item $H = \partial_tV_\Sigma/(3V_\Sigma)$ is the expansion rate of $\Sigma$ with $V_\Sigma(t)$ its volume.
        \item $\T\Xi$ is a traceless (i.e. ${\Xi_c}^c \coloneqq 0$) and transverse (i.e. $D_c{\Xi_a}^c \coloneqq 0$) tensor.
        \item $\T{a}_{\not= \rm grav}$ is the non-gravitational 3-acceleration acting on the fluid.
        \item $\rho$ is the mass density of the fluid, and $M_{\rm tot}$ is the total mass in $\Sigma$.
        \item The operator $\ \widehat{}\ $ acts on a scalar $\psi$ as $\widehat{\psi} \coloneqq \psi - \Saverage{\psi}$, with $\Saverage{\psi}(t) \coloneqq \frac{1}{V_\Sigma}\int_\Sigma \psi \sqrt{\mathrm{det}(h_{ab})}\dd^3 x$ being the average over the whole volume of $\Sigma$.
\end{enumerate}
The gravitational system~\eqref{eq::NEN_def_g_1}--\eqref{eq::NEN_sys_av_rho} is algebraically equivalent to the gravitational system in classical Newton's theory with the presence of an anisotropic expansion \cite{1997_Buchert_et_al, 2021_Vigneron}. The only difference is the spatial Ricci tensor $\T\CR$, relative to the spatial metric $\T h$ and its connection $\T D$, which is not zero but given by formula~\eqref{eq::NEN_sys_Ri}. This curvature tensor allows these equations to be defined on either a spherical or hyperbolic topology, depending on the sign of the scalar curvature $\CR$. If we assume $\T\CR$ to be zero, then we retrieve Newton's equations exactly.

Equation~\eqref{eq::NEN_def_g_1} corresponds to Newton's second law for the spatial acceleration of the fluid spatial velocity (this is also the Navier--Stokes equation, since it is written for a fluid), with $\left(\partial_t - \Lie{\T \beta}\right) v^a + v^c\D_c v^a$ being the spatial acceleration of the spatial velocity $\T v$ in any coordinate system; Eq.~\eqref{eq::NEN_New_Gaus_1} is the cosmological Poisson equation; Eq.~\eqref{eq::NEN_g_2} constrains the gravitational field to be irrotational (i.e. no gravitomagnetism); Eq.~\eqref{eq::NEN_Theta_hb} is the evolution equation for the spatial metric; Eq.~\eqref{eq::NEN_cont} is the continuity equation; and Eq.~\eqref{eq::NEN_sys_av_rho} is the expansion law for the volume of the manifold $\Sigma$, and corresponds to the Friedmann law in the case of isotropic expansion.

$\T\Xi$ is a transverse shear and is also present in Newton's theory \cite{1955_Heckmann_et_al, 1956_Heckmann_et_al, 1997_Buchert_et_al, 2021_Vigneron}, where it models an anisotropic expansion. If this term is assumed to be zero, then there exists a coordinate system, i.e. a choice of $\T \beta$, in which the spatial metric takes the simple form $h_{ab} = a^2(t)\tilde{h}_{ab}(x^i)$ with $\dot{a}/a = H$ and where the Ricci tensor $\T{\tilde{\CR}}$ associated to $\T{\tilde{h}}$ is ${\tilde\CR}_{ab} = {\initial{\CR}}/3 \, \tilde{h}_{ab} = {\CR}_{ab} $, where the subscript $\initial{}$ stands for initial. The coordinate system implying this form of the spatial metric corresponds to $\T\beta = -\T v$ and is called an \textit{inertial coordinate system}, or Eulerian coordinate system in the language of fluid dynamics. In this system, the spatial metric has no local dynamics as it is separated in space and time.

\subsection{Simplified form of the gravitational system in the NEN theory}
\label{sec:NEN_grav_simp}

Hereafter, we set $\T\beta = - \T v$ and $\T\Xi = 0$. This latter choice, in addition to implying a separation between the space and time dependence of the spatial metric, is also in agreement with the fact that there are no strong observational claims of a global anisotropy in the expansion of our Universe. Thus, we end up with the following simplified gravitational system, where we introduce the gravitational potential $\Phi$, defined by $\T g = - \T{D}\Phi$:
\begin{align}
        \left(\partial_t - v^cD_c\right) v^a+ 2v^a H  &= -h^{ac}D_c \Phi + (a_{\not= \rm grav})^a, \label{eq::NEN_def_g_2} \\
        \left(\partial_t - v^cD_c\right) \rho &= -\rho\left(3H + \D_c v^c\right), \label{eq::NEN_cont_2} \\
        h^{cd}D_cD_d \Phi &= 4\pi G \widehat{\rho}, \label{eq::NEN_Poisson}
\end{align}
where $\widehat{\rho}$ \ is the density deviation (defined in Sec.~\ref{sec:NEN_grav_gene}), and the expansion law
\begin{align}
        3\left(\dot H + H^2\right) + 4\pi G \frac{M_{\rm tot}}{V_{\Sigma}(t)} - \Lambda = 0, \label{eq::NEN_exp_2}
\end{align}
where we have in spherical coordinates $(\xi,\theta,\varphi)$
\begin{align}
        h_{ab} =  \frac{6 \, a^2(t)}{\initial{\CR}}\textrm{diag}\left[1, \sinn^2\xi, \sinn^2\xi\sin^2\theta\right]_{ab}, \label{eq::h_ab_sphe}
\end{align}
with $a$ being dimensionless and $\dot a/a = H$ and
\begin{align}
{\sinn \, \xi\ \coloneqq}
\begin{cases}
        \sinh \xi,\quad &\text{if } \initial{\CR} < 0 \text{ (hyperbolic)} \\
        \sin \xi, \quad &\text{if } \initial{\CR} > 0 \text{ (spherical)}.
\end{cases}
\label{eq::sinn}
\end{align}
Once again, this system of equations is equivalent to the cosmological Newton equations, but with the (implicit) presence of spatial curvature (i.e. $\CR_{ij} = \CR/3 \, h_{ij}$) in the spatial derivative. In particular, as in cosmology based on Newton's equations, the density in the Poisson equation arises as the difference from the average density on $\Sigma$, i.e. it is the density deviation $\widehat{\rho}$ rather than the absolute density $\rho$. This is the main difference with respect to the NEN theory proposed by \cite{2009_Roukema_et_al, 2020_Barrow}, who used the absolute density that led to a white hole (see Sec.~3 in \cite{2022_Vigneron_b}).

From Eq.~\eqref{eq::NEN_exp_2}, we see that the expansion law is the same as in Newton's theory, which corresponds to Friedmann's expansion law, and this holds for any inhomogeneous solution for $\rho$ and $\T v$. This implies that there are no effects of the inhomogeneities in the global expansion (such an effect is often called the cosmological backreaction in general relativity), no matter the class of topology chosen (here Euclidean, spherical or hyperbolic), thus answering the question raised in \cite{2022_Vigneron}. Therefore, the only difference with (Euclidean) Newton's theory that might come from a non-Euclidean topology will be a local influence on structure formation, coming either from spatial curvature or the precise choice of topology, i.e. the choice of multiconnexity. A full study of these effects requires a $N$-body simulation with the above system of equations. A $N$-body simulation requires knowledge of the gravitational potential related to a single point mass obtained as a solution of the Poisson Eq.~\eqref{eq::NEN_Poisson} with a Dirac $\delta$ field for $\rho$. In the following we present this solution in the case where $\Sigma$ is a spherical manifold and as a function of the (multiconnected) topology, i.e. as a function of the shape of the Universe.

\section{Gravitational potential in the regular spherical topologies}
\label{sec:pot}

\begin{table*}[t]
	\newcommand\namecolwidth{0.55}
	\renewcommand{\arraystretch}{1.5}
	\captionsetup{margin={0pt,50pt}}
	\centering
	\caption{List of the eight orientable spherical topologies definable with a Platonic solid fundamental domain (FD) (see Tables~1 and~3 of \cite{2004_Everitt} and Table~1 of\cite{2009_Cavicchioli_et_al_a}).
    Columns indicate the 3-manifold name $\Sigma$ defined in \cite[][Table~1]{2009_Cavicchioli_et_al_a}; the shape of the initial choice of FD for defining the space; other names; whether or not the space is guaranteed to be globally homogeneous by being a single-action spherical 3-manifold (Sec.~4.1 in \cite{2001_Gausmann_et_al}); the number $N_\Sigma$ of copies of the FD that tile $\mS^3$. \label{tab:Cavicchioli_M1_M8}}
	\begin{tabular}{lp{3cm}p{9.cm}cr}
		\hline\hline
		\multicolumn{1}{c}{Space $\Sigma$} & \multicolumn{1}{c}{ Initial FD} & \multicolumn{1}{c}{Names} & \multicolumn{1}{c}{Single action} & \multicolumn{1}{c}{$N_\Sigma$} \\
		\hline
		$M_1$ & Tetrahedron & $L(5,3)$ \protect\cite{2009_Cavicchioli_et_al_a} & no & 5 \\
		$M_2$ &  Cube & $L(8,3)$ \protect\cite{2009_Cavicchioli_et_al_a} & no & 8 \\
		$M_3$ &  Cube & Quaternion space, 4-sided prism space,  $S^3/_{D_2^*}$ (Sect.~4.1 in \cite{2001_Gausmann_et_al}) & \textbf{yes} & 8  \\
		$M_4$ &  Octahedron & $S^3/_{Q_8 \times \mathbb{Z}_3}$ (Table~4 in \cite{2009_Cavicchioli_et_al_a}) & no & 24\\
		$M_5$ &  Octahedron & $S^3/_{D_{24}}$ (Table~4 in \cite{2009_Cavicchioli_et_al_a}) & no & 24 \\
		$M_6$ &  Octahedron & Octahedral space,
		$S^3/_{T^*}$ (Sect.~4.1 in \cite{2001_Gausmann_et_al}) & \textbf{yes}  & 24 \\
		$M_7$  &  Dodecahedron & Poincar\'e homology 3-sphere, Poincar\'e dodecahedral \newline  \textcolor{white}{f\quad} space, $S^3/_{I^*}$ (Sect.~4.1 in \cite{2001_Gausmann_et_al}) & \textbf{yes} & 120 \\
		$M_8$ &  Dodecahedron & $S^3/_{P_{24} \times \mathbb{Z}_5}$ (Table~4 in \cite{2009_Cavicchioli_et_al_a}) & no & 120 \\
		\hline\hline
	\end{tabular}
\end{table*}

\subsection{Topological terminology}
\label{sec:top_terminology}

For an introduction to topology-related terminology in the context of cosmic topology, see \cite{1995_La_Lu}, and for the spherical case, see \cite{2001_Gausmann_et_al}.
Key terms include the 3-manifold itself $\Sigma$ (here referred to loosely as ``a topology'', to focus on topological properties); the covering space $\widetilde{\Sigma}$ (which in the case of interest here will be $\widetilde{\Sigma} = \mS^3$); and the group $\Gamma$ of holonomies (a particular type of smooth mapping from $\widetilde{\Sigma}$ to itself) that relates $\Sigma$ and $\tilde\Sigma$ via $\Sigma = \widetilde{\Sigma}/\Gamma$. Applying every mapping $\gamma_i$ that is a member of $\Gamma$ to a single ``tile'' -- a fundamental domain (FD, filled-in polyhedron in this case) of $\Sigma$ -- gives a full tiling of, in our case, the 3-sphere.
We use the index 0 for the identity holonomy: $\gamma_0(x) = x, \forall x \in \Sigma$.
The fundamental domain shape of $\Sigma$ is not, in general, unique -- for instance the Klein bottle (a 2-dimensional manifold) can be tiled by either a hexagon or a rectangle.

A topology is said to be \textit{globally homogeneous} if the distance between a test particle and its image in a neighbouring tile (within the covering space) is independent of the particle's position. For a more formal definition of global homogeneity and the role of Clifford translations (see Sec.~4.1 in \cite[][]{2001_Gausmann_et_al}). We describe a topology as \textit{regular} if its fundamental domain is unique (a consequence of global homogeneity) and is a platonic solid.

\subsection{Choice of topologies}
\label{sec:sphe_top}

There are an infinite number of multiply connected spherical and hyperbolic topologies. In practice, studies of cosmic topology, whether observational or theoretical, usually only consider a small number of topologies, in particular, those for which the fundamental domain is unique and is a Platonic solid\footnote{Some examples of ``non-Platonic'' topologies studied in cosmology are the duct space \cite{2000_Roukema}, the slab space \cite{2015_Planck_XVIII}, and the truncated cube space \cite{2004_Weeks_et_al, 2009_Roukema_et_al}, which we will not consider in this paper.}. The main reason behind this choice is that these topologies are globally homogeneous and regular, which follows the spirit of the cosmological principle. In the present paper, we will only focus on the spherical ``Platonic'' topologies, leaving the hyperbolic ones for a later study. The reason for this is that solving the NEN equations is highly non-trivial in the multiply connected hyperbolic case in comparison with the spherical case (see the remark in the Appendix). Furthermore, several recent studies of cosmological data that infer a non-zero spatial curvature favour a spherical topology (e.g. \citep{2020_Di-Valentino_et_al, 2021_Di-Valentino_et_al_bis, 2021_Handley}).

Among 3-dimensional spaces, there are exactly eight orientable spherical topologies that can be defined starting from a Platonic solid as a fundamental domain (see Tables~1 and~3 in \cite{2004_Everitt}).
These are labelled $M_1, \ldots, M_8$ (see Table~1 in \cite{2009_Cavicchioli_et_al_a}),
and are listed in Table~\ref{tab:Cavicchioli_M1_M8} of the present paper. Two of these (spaces $M_1$ and $M_2$) can be equivalently constructed using a lens fundamental domain (see Sec.~4 in \citep{2001_Gausmann_et_al})\footnote{Section~3 of \cite{2009_Cavicchioli_et_al_a} shows that $M_1$ is equivalent to $L(5,3)$ and $M_2$ is equivalent to $L(8,3)$, where $L(p,q)$, for $p,q$ coprime and $0<q<p$, means that $p$ copies of the lens fundamental domain of central thickness $2\pi /p$ fill the 3-sphere, each matched after a rotation of $2\pi q/p$}. Thus, the fundamental domain is not unique in these two cases, which are therefore not globally homogeneous.

Among the six remaining spaces, $M_3, \ldots, M_8$, the fundamental domain used for the construction is either the cube, the octahedron, or the dodecahedron. Spherical spaces that are single-action spherical 3-manifolds are necessarily globally homogeneous (see Sec.~4.1 in \cite{2001_Gausmann_et_al}). In conclusion, from Table~\ref{tab:Cavicchioli_M1_M8}, the topologies of interest for this paper are the spaces $M_3$, $M_6,$ and $M_7$. They, respectively, correspond to a regular tiling of the 3-sphere by 8, 24, and 120 copies of the FD.\saut

\remark{The tilings of the 3-sphere by 5, 16, or 600 black holes considered in lattice cosmology \cite[e.g.][]{2013_Clifton_et_al} do not correspond to either a globally homogeneous topology, or to a topological space, and therefore are not of interest for the present paper.}

\subsection{Gravitational potential of a point mass in spherical topologies}
\label{sec:grav_sys_pm}

In this paper we want to calculate the gravitational potential $\Phi_\Sigma$ of a point mass $M$ in the spherical topologies chosen above. This potential is required to be able to perform $N$-body simulations in these topologies, which is one of the goals of the NEN theory.

We consider the mass to be at rest and for simplicity (without loss of generality) we assume that it is placed at the north pole of the 3-sphere with initial curvature radius 1, i.e. $\initial{\CR} = 6$. Therefore, the density is described by the Dirac field $\rho = M\delta_\Sigma^{(0,0,0)}(\xi,\theta,\varphi)$, centered on the coordinates $(0,0,0)$, of the Riemannian manifold $(\Sigma, \T h)$; and the average density is $\Saverage{\rho} = M/V_\Sigma$. The Poisson Eq.~\eqref{eq::NEN_Poisson}, constraining the gravitational potential $\Phi_\Sigma$ created by the mass $M$, becomes
\begin{align}
        &h^{cd}D_cD_d\Phi_\Sigma = 4\pi GM\left(\delta_\Sigma^{(0,0,0)}(\xi,\theta,\varphi) - \frac{1}{V_\Sigma} \right), \label{eq:Poisson_M}
\end{align}
where the spatial metric in spherical coordinates is given by~\eqref{eq::h_ab_sphe}.

To solve~\eqref{eq:Poisson_M} we use the same method as in \cite{2009_Roukema_et_al} by splitting the equation over all the images of $M$ in the covering space $\mS^3$. We have
\begin{align}
   \delta_\Sigma^{(0,0,0)}(\xi,\theta,\varphi) = \sum_{\gamma_{i} \in \Gamma} \delta_{\mS^3}^{\gamma_i(\,(0,0,0)\,)} (\xi,\theta,\varphi),
\end{align}
where $\delta_{\mS^3}^{\gamma_i(\,(0,0,0)\,)}$ denotes the Dirac field of $\mS^3$ centered at the holonomy position $\gamma_i(\,(0,0,0)\,)$. This split is well defined because there are a finite number of images on $\mS^3$. This is not the case for the 3-torus \citep{2016_Steiner} or hyperbolic topologies (see the remark in Appendix~\ref{sec:disc_H3}). Using $V_\Sigma = V_{\mS^3}/N_\Sigma$, where $N_\Sigma$ is the number of images of the fundamental domain of $\Sigma$ on $\mS^3$, we have
\begin{align}
        \delta_\Sigma^{(0,0,0)}(\xi,\theta,\varphi) - \frac{1}{V_\Sigma} = \sum_{\gamma_{i} \in \Gamma}\left(\delta_{\mS^3}^{\gamma_i(\,(0,0,0)\,)}(\xi,\theta,\varphi) - \frac{1}{V_{\mS^3}} \right). \label{eq:splitting}
\end{align}
By linearity of the Laplacian and because the above sum is finite, we can write $\Phi_\Sigma$ in the form
\begin{align}
  \Phi_\Sigma = \sum_{\gamma_{i} \in \Gamma} \Phi_{\mS^3} ^{\gamma_i(\,(0,0,0)\,)}, \label{eq:Phi_Sigma_sum}
\end{align}
with each $\Phi_{\mS^3} ^{\gamma_i(\,(0,0,0)\,)}$ solution of the equation
\begin{align}
        h^{cd}D_cD_d \Phi_{\mS^3} ^{\gamma_i((0,0,0))} = 4\pi GM\left(\delta_{\mS^3}^{\gamma_i(\,(0,0,0)\,)}(\xi,\theta,\varphi) - \frac{1}{V_{\mS^3}} \right).
\end{align}
In other words, the gravitational potential of one point mass in $\Sigma$ corresponds to the sum of the potential (as calculated in $\mS^3$) of all the point mass images on $\mS^3$. Thus, it is sufficient to solve the Poisson equation for a single generic image in $\mS^3$; i.e. we need to solve
\begin{align}
        \partial_\xi^2 \Phi_{\mS^3}^{(0,0,0)} + 2\cot\xi \, \partial_\xi \Phi_{\mS^3}^{(0,0,0)} = 4\pi GM\left(a^2\delta(\xi) - \frac{1}{2\pi^2 a}\right).
\end{align}
The solution is
\begin{align}
        \Phi_{\mS^3}^{(0,0,0)}(t,\xi,\theta,\varphi) = -\frac{GM}{a}\left[(\cot\xi)\left(1-\xi/\pi\right) + A\right], \label{eq:phi_sol_S3}
\end{align}
where $A$ is an integration constant which is not physical and sets the convention we want to take for the value of $\Phi_{\mS^3}^{(0,0,0)}$ at the south pole ($\xi = \pi$). Imposing $\lim_{\xi \rightarrow \pi} \Phi_{\mS^3}^{(0,0,0)}(t, \xi) = 0$ corresponds to $A = 1/\pi$.

We see that the potential~\eqref{eq:phi_sol_S3} depends only on $\xi$, which is the comoving distance between the image at $(0,0,0)$ and the point at coordinates $(\xi,\theta,\varphi)$. Therefore, any term $\Phi_{\mS^3} ^{\gamma_i(\,(0,0,0)\,)}(\xi,\theta,\varphi)$ entering in the sum~\eqref{eq:Phi_Sigma_sum} can be obtained by replacing $\xi$ in~\eqref{eq:phi_sol_S3} with the comoving distance $d_{(i)} (\xi,\theta,\varphi)$ between the point at coordinates $(\xi,\theta,\varphi)$ and the $i^{\rm th}$ image at coordinates $\gamma_i(\,(0,0,0)\,)$. Equation~\eqref{eq:Phi_Sigma_sum} becomes
\begin{align}
        &\label{eq:Phi_app} \Phi_\Sigma(t,\xi,\theta,\varphi) = \\
        &-\frac{GM}{a(t)}\sum_{\gamma_{i} \in \Gamma} \left[\cot \left(d_{(i)}(\xi,\theta,\varphi)\right) \left(1- \frac{d_{(i)}(\xi,\theta,\varphi)}{\pi}\right) + A\right]. \nonumber
\end{align}
This is the gravitational potential created by one mass point in \textit{any} spherical topology.\\

\remark{We only consider globally homogeneous topologies, so changing the position of the point mass only changes the gravitational field by a 4-dimensional rotation on the 3-sphere. This is not the case for an inhomogeneous topology. Applying the full group of holonomies $\Gamma$ to the point mass in an inhomogeneous topology yields a set of images whose distribution varies depending on the position of the point mass in $\Sigma$.}

\subsection{Embedding method for calculation}

In view of calculating the Taylor series of formula~\eqref{eq:Phi_app}, it can be cumbersome to keep the 3-dimensional coordinates $(\xi, \theta, \varphi)$. A simpler method, allowing in particular for an easier computation of the distances $d_{(i)}(\xi,\theta,\varphi)$, is to use an embedding of $\mS^3$ in $\mE^4$ so that the metric on $\mS^3$ is preserved (as in \cite{2009_Roukema_et_al}): a point $(\xi, \theta, \varphi)$ on the 3-sphere is described by a 4-vector $\T X$ in $\mE^4$ such that $X_\mu X^\mu = 1$, where Greek indices run from $0$ to $3$\footnote{The pull-back of the flat metric of $\mE^4$ on the hypersurface defined by $X_\mu X^\mu = 1$ is a 3-metric of constant scalar curvature. This means that this embedding of the 3-sphere in $\mE^4$ preserves the spatial metric.}. The embedding is not physical but just a mathematical trick to simplify the calculation of the $d_{(i)}$. The mapping onto the 3-sphere, i.e. from $\{X^\mu\}_{\mu=0,1,2,3}$ to $(\xi,\theta,\varphi)$, is made with {hyperspherical~coordinates}:
\begin{align}
\begin{cases}{\label{eq::coord_sph}}
                X^0 = \cos{\xi} \\
                X^1 = \sin{\xi}\sin{\theta}\cos{\varphi} \\
                X^2 = \sin{\xi}\sin{\theta}\sin{\varphi} \\
                X^3 = \sin{\xi}\cos{\theta}
\end{cases},
\end{align}
and the distance $\dd\left[\T X, \T Y\right]$, on the 3-sphere, between two points $X^\mu$ and $Y^\mu$ is given by
\begin{equation}
        \dd\left[\T X, \T Y\right] = \arccos\left({\T X \cdot \T Y}\right).
\end{equation}
Thus, the calculation of distances on $\mS^3$ corresponds to the calculation of a scalar product in $\mE^4$ (introduced in observational cosmology for the spherical and hyperbolic cases in \cite{2001_Roukema}). Setting the positions of the topological images of the point mass as unit 4-vectors $\T Y_{(i)}$, we obtain $d_{(i)} = \arccos\left({\T X
\cdot \T Y_{(i)}}\right)$, with $\T X$ defined by \eqref{eq::coord_sph}. Then, using the trigonometric relation $\cot\arccos x = x/\sqrt{1-x^2}$, the gravitational potential~\eqref{eq:Phi_app} becomes
\begin{align}
        &\frac{a(t)}{GM}\Phi_\Sigma(t,\T X) = -N_\Sigma A \label{eq:phi_sol} \\
        &\qquad - \sum_{{\{\T Y_i\, :\, \gamma_i \in \Gamma  \}}}  \frac{\T X \cdot \T
Y_{(i)}\, \left(1- \arccos\left(\T X \cdot \T Y_{(i)}\right)/\pi\right)}{\sqrt{1- \left(\T X \cdot \T Y_{(i)}\right)^2}}. \nonumber
\end{align}

For the specific cases considered in this paper (spaces $M_3$, $M_6$ and $M_7$ in Table~\ref{tab:Cavicchioli_M1_M8}), the positions $\T{Y}_i := \gamma_i(\,(0,0,0)\,)$ can be found in \cite{2012_Clifton_et_al} (Table~3) for $M_3$ (eight terms in the sum), and in \cite{2001_Gausmann_et_al} (Appendix~B) for $M_6$ (24 terms) and $M_7$ (120 terms). The results of Table~\ref{tab:results} are derived from formula~\eqref{eq:phi_sol} with these positions.\\

\newpage
\begin{widetext}
\subsection{$N$-body description}
\label{sec:N-body}

Formula~\eqref{eq:Phi_app} is the gravitational potential induced by one point mass in $\Sigma$. For a distribution of $N$ masses $M_n$ at positions $(\xi_n(t), \theta_n(t), \varphi_n(t))$, the total gravitational potential at $(\xi,\theta,\varphi)$ is given by

\begin{align}
        \Phi_{\rm tot}(t,\xi,\theta,\varphi) = -\frac{GM_{\rm tot}}{a(t)} N_\Sigma A - \sum^N_{n=1}\left[\frac{GM_n}{a(t)} \sum_{{\{ \gamma_i \in \Gamma  \}}} \cot \left(d_{(n,i)}(t,\xi,\theta,\varphi)\right) \left(1- \frac{d_{(n,i)}(t,\xi,\theta,\varphi)}{\pi}\right)\right], \label{eq:Phi_N_tot}
\end{align}

where $d_{(n,i)}(t,\xi,\theta,\varphi)$ is the distance between the point at $(\xi,\theta,\varphi)$ and the i$^{\rm th}$ image of the mass $M_n$. We recall that the expansion law describing the evolution of the scale factor $a(t)$ is given by Eq.~\eqref{eq::NEN_exp_2} and does not depend on local dynamics.

Using the embedding in $\mE^4$ and denoting by $\T Y_{(n)}(t)$ the position of $M_n$ (inducing a set of images at positions $\T Y_{(n,i)}(t)$), formula~\eqref{eq:Phi_N_tot} can be rewritten

\begin{align}
        \Phi_{\rm tot}(t,\T X) = -\frac{GM_{\rm tot}}{a(t)} N_\Sigma A - \sum^N_{n=1}\left[\frac{GM_n}{a(t)} \sum_{{\{\T Y_i\, :\, \gamma_i \in \Gamma  \}}}  \T X \cdot \T Y_{(n,i)}(t)\frac{ \left(1- \arccos\left(\T X \cdot \T Y_{(n,i)}(t)\right)/\pi\right)}{\sqrt{1- \left(\T X \cdot \T Y_{(n,i)}(t)\right)^2}}\right]. \label{eq:Phi_N_tot_4D}
\end{align}

As with Newton's theory, the dynamics of the point masses is then solved via Newton's second law~\eqref{eq::NEN_def_g_2}, which writes for each mass, denoting $x^i_{(n)}(t) = (\xi_n(t), \theta_n(t), \varphi_n(t))$,
\begin{align}
	\ddot{x}^i_{(n)} + \Gamma^i_{cd}(x^k_{(n)})\, \dot x^c_{(n)} \dot x^d_{(n)} &+ 2H\dot{x}^i_{(n)} =  \label{eq:2nd_law_N_tot} -\left(h^{ij}\partial_j \Phi_{\rm tot}\right)(t, x^k_{(n)}),
\end{align}
where $\Gamma^i_{kl}(x^k_{(n)})$ are the Levi-Civita coefficients at position $x^k_{(n)}$ of the metric~\eqref{eq::h_ab_sphe}.
\end{widetext}

Relations~\eqref{eq:Phi_N_tot} and \eqref{eq:2nd_law_N_tot} are valid for $N$-body simulations of non-relativistic structure formation in any spherical topology. For performing such simulations, the quaternion space $M_3$ is likely to be the easiest to implement numerically (see also \citep[][5.4]{2021_Vigneron_PhD}). The advantage of $M_3$ is that the fundamental domain of this topology is a cube, as in the case of the 3-torus usually used in cosmological simulations. However, the $M_3$ case differs in that curvature is positive rather than zero and faces are identified differently than in the 3-torus case: each holonomy is a screw motion, corresponding to a translation by a fundamental domain length and a turn of $\pi/2$. Implementation of $M_6$ and $M_7$ will be more difficult, but is likely to be needed (especially the case of $M_7$) to study how separable the effects of curvature and topology are (see Sec.~\ref{sec:dominant}).

While the exact solution~\eqref{eq:Phi_app} (which is generalised to $N$ bodies in ~\eqref{eq:Phi_N_tot}) is needed for (analytically) exact $N$-body simulations, the calculations of its Taylor series can give interesting information on the effects that topology and curvature may have on structure formation. This Taylor series is derived in the next section up to fifth order. We also discuss different conventions for the choice of the (non-physical) integration constant $A$.

\begin{table*}[t]
\captionsetup{margin={0pt,40pt}}
\centering
\caption{Taylor expansion series of the gravitational potential ($GM = 1$) near a point mass in the infinite flat space and the 3-torus (given by, e.g., formula~(4.24) in \cite{2016_Steiner}); in all the regular spherical topologies [Eq.~\eqref{eq:Phi_app}], as a function of $\CR$ and $V_\Sigma$ when they are non-zero; and for the simply connected hyperbolic case as given below in \eqref{eq:Phi_hyperbolic}. Anisotropic terms, i.e. featuring a dependence on $\theta$ or $\varphi$, are not shown and left as ``anis.''. The two natural conventions for the zeroth order $\Phi_0$ are shown in Tables~\ref{tab:results_A} and~\ref{tab:results_Phi_0}. The rows with entries for $N_\Sigma$ refer to quotients of $\mS^3$ (Table~\protect\ref{tab:Cavicchioli_M1_M8}). The case of a point mass in the hyperbolic space $\mH^3$ is also presented, to support the interpretations of these results made in Sec.~\ref{sec:disc_even_odd}, in this case $\CR = -6$. However, $\mH^3$ has the same topology as $\mE^3$, so these are the same topological 3-manifold, but with different curvatures (see the discussion in the Appendix concerning the physical relevance of this solution).}
\label{tab:results}
	\renewcommand{\arraystretch}{1.3}
\begin{tabular}{l x{2cm}x{2cm} x{2cm}x{2cm}x{2cm}x{2cm}x{2cm}}
\hline\hline  
Topology & $N_\Sigma$ & $\Phi_{-1}$ & $\Phi_1$ & $\Phi_2$  & $\Phi_3$ & $\Phi_4$ & $\Phi_5$ \\
\hline
\multicolumn{8}{c}{Euclidean (infinite or Thurston-type)}  \\
$\mE^3$ &  & $-1$ & 0 & 0 & 0 & 0 & 0 \\
$\mathbb{T}^3$ &  & $-1$ & 0 & $- \frac{2\pi}{3}\frac{1}{{V}_\Sigma}$ & 0 & anis. & $\ 0$
\rule[-1.4ex]{0ex}{0.1ex}\vspace{.2cm}\\

\multicolumn{8}{c}{Spherical}  \\ \vspace{.1cm}
$\mS^3$ & 1 & $-1$ & $\frac{1}{3}\frac{\CR}{6}$ & $- \frac{2\pi}{3}\frac{1}{{V}_\Sigma}$ & $\frac{1}{45}\left(\frac{\CR}{6}\right)^2 $ & $- \frac{2\pi}{45}\frac{\CR/6}{{V}_\Sigma}$ & $\frac{2}{945}\left(\frac{\CR}{6}\right)^3$ \\
$M_3$ & 8 & $-1$ & $\frac{1}{3}\frac{\CR}{6}$ & $- \frac{2\pi}{3}\frac{1}{{V}_\Sigma}$ & $\frac{1}{45}\left(\frac{\CR}{6}\right)^2 $ & anis. & $\frac{2}{945}\left(\frac{\CR}{6}\right)^3$ \\
$M_6$ & 24 & $-1$ & $\frac{1}{3}\frac{\CR}{6}$ & $- \frac{2\pi}{3}\frac{1}{{V}_\Sigma}$ & $\frac{1}{45}\left(\frac{\CR}{6}\right)^2 $ & anis. & $\frac{2}{945}\left(\frac{\CR}{6}\right)^3$ \\
$M_7$ & 120 & $-1$ & $\frac{1}{3}\frac{\CR}{6}$ & $- \frac{2\pi}{3}\frac{1}{{V}_\Sigma}$ & $\frac{1}{45}\left(\frac{\CR}{6}\right)^2 $ & $- \frac{2\pi}{45}\frac{\CR/6}{{V}_\Sigma}$ & $\frac{2}{945}\left(\frac{\CR}{6}\right)^3$
\rule[-1.4ex]{0ex}{0.1ex}\vspace{.2cm}\\

\multicolumn{8}{c}{Hyperbolic (infinite)}  \\ \vspace{.1cm}
$\mH^3$ &  & $-1$ & $\frac{1}{3}\frac{\CR}{6}$ & 0 & $\frac{1}{45}\left(\frac{\CR}{6}\right)^2 $ & 0 & $\frac{2}{945}\left(\frac{\CR}{6}\right)^3$ \rule[-1.4ex]{0ex}{2.6ex} \\
\hline\hline
\end{tabular}
\end{table*}

\subsection{Leading order solutions}
\label{sec::sol}

In this section we analyse the form of the potential close to the location of the point mass. We calculate the Taylor expansion series of the potential as a function of the physical distance $r = a\xi$ to that point mass, where we define the different orders $\Phi_n$ with
\begin{align}
  \frac{1}{GM}\Phi_\Sigma(t,\T X) = \sum_{n=-1}^{\infty} \Phi_n(r, \theta, \varphi) \, r^n.
      \label{eq:phi_expansion_definition}
\end{align}
We provide the results for each topology in Table~\ref{tab:results}, obtained using the software {\sc Maxima} (see ``Data and code availability'' at the end of the paper). We only give the orders that remain isotropic (i.e. depend only on $r$) and write them as a function of the volume ${V}_\Sigma := 2\pi^2a^3/N_\Sigma$ of the manifold (i.e. volume of the fundamental domain, if defined) and its curvature $\CR$ (if non-zero). We also provide the solution in the case in which the manifold is $\mE^3$, $\mT^3$ (with $V_{\mT^3} = a^3$), or $\mH^3$. The zeroth order $\Phi_0$ is not shown as it depends on the value of $A$ present in the sum~\eqref{eq:phi_sol}. We stress that $\Phi_0 \not= -N_\Sigma A$.

Two natural conventions are possible for setting the value of the constant $A$ as a function of the topology.
\begin{enumerate}
        \item \label{caca} Require $\Phi_0 = 0$. This is similar to requiring the vanishing of the potential at infinity in $\mE^3$. The values of $A$ for this convention are given in Table~\ref{tab:results_A}.  Using formula~\eqref{eq:phi_sol} with these values in $N$-body numerical simulations would avoid the need to calculate the zeroth order for each particle, which might increase numerical efficiency.
        \item \label{bite} Require the average of the potential over the volume of $\Sigma$ to be zero:
\begin{align}
        \int_{V_\Sigma} \Phi_\Sigma(t,\T X) \, a^3\sin^2{\xi}\sin\theta\, \dd \xi\, \dd\theta\, \dd\varphi = 0. \label{eq:conv_plasma}
\end{align}
In this convention, adopted in crystallography and plasma physics for $\mT^3$, we have $\Phi_0 \not=0$. The value of $\Phi_0$ in this case (called the Madelung constant) is generally interpreted as the total interaction energy created by one particle in $\Sigma$ (see e.g. \citep{1966_Brush_et_al}). However, it is unclear if this interpretation is meaningful in the case of spherical topologies. The values of  $\Phi_0 \, V_\Sigma^{1/3}$ (i.e, scaled to be adimensional at a fixed volume) in this convention are provided for completeness in Table~\ref{tab:results_Phi_0}. These are obtained by choosing $A = -1/(2\pi)$ for each topology.
\end{enumerate}
While we do not expect the value of $A$ for convention (i) to have physical significance, the value of $\Phi_0$ in convention (ii) could be interpreted physically, as is the case in the relation between crystallography and the 3-torus \citep{1966_Brush_et_al}.

\begin{table}[b]
\centering
\begin{tabular}{lx{1cm}x{4.2cm}c}
\hline\hline
Topology & $N_\Sigma$ & \multicolumn{2}{c}{Integration constant $A$ for $\Phi_0 = 0$}\vspace{.15cm}\\
  && Analytical & Numerical \\
\hline\vspace{-.1cm}\\
$\mS^3$ & 1 & $ \frac{1}{\pi}$ & $0.3183$
\vspace{.2cm}\\
$M_3$ & 8 & $\frac{1}{4\pi}$ & $0.0796$
\vspace{.2cm}\\
$M_6$ & 24 &
$ \frac{\left(9 - 4\sqrt{3}\pi\right)}{108\pi}$ & $-0.0376$
\vspace{.2cm}\\
$M_7$ & 120 &
$\frac{1}{60\pi} - \frac{1}{50} \sqrt{10+\frac{22}{\sqrt{5}}} - \frac{1}{18 \sqrt{3}}$ & $-0.1159$
\vspace{.2cm}\\
\hline\hline
\end{tabular}
\caption{Values of the constant $A$ in formula~\protect\eqref{eq:phi_sol} in case (\ref{caca}), where we impose the convention $\Phi_0 = 0$, as a function of the regular spherical topologies.
  \label{tab:results_A}}
\end{table}

\begin{table}[t]
\centering
\begin{tabular}{lx{1.9cm}x{4.8cm}}
\hline\hline
Topology & $N_\Sigma$ & $\Phi_0 \, V_\Sigma^{1/3}$ for $\int_{V_\Sigma} \Phi \, \dd V = 0$\\
\hline
\multicolumn{3}{c}{ Euclidean} \\
$\mathbb{T}^3$ &  & 2.837
\rule[-1.4ex]{0ex}{0.1ex}\\
\multicolumn{3}{c}{Spherical} \\
$\mS^3$ & 1 & $1.290$
\vspace{.2cm}\\
$M_3$ & 8 & $2.581$
\vspace{.2cm}\\
$M_6$ & 24 & $2.733$
\vspace{.2cm}\\
$M_7$ & 120 & $2.847$
\vspace{.2cm}\\
\hline\hline
\end{tabular}
\caption{Values of $\Phi_0 \, V_\Sigma^{1/3}$ (adimensional value at fixed volume) in case (\ref{bite}), in which we impose the integral convention~\eqref{eq:conv_plasma}, as a function of the regular spherical topologies. These values are obtained with the choice $A = -1/(2\pi)$, which cancels the average of the potential for each topology. We also provide the value in the case of $\mT^3$ given by the zeroth order of the Ewald summation \citep{1966_Brush_et_al,2016_Steiner}. \label{tab:results_Phi_0}}
\end{table}

\subsection{Consistency check}

To be coherent with the non-relativistic theory in Euclidean topologies, i.e. Newton's theory, we should retrieve Newton's law from the limit $\CR \rightarrow 0^+$ of the $\mS^3$ solution~\eqref{eq:phi_sol_S3}. For this, we reintroduce $\CR$ and write the solution as function of the physical distance $r = \xi/\sqrt{\CR/6}$:
\begin{align}
        \Phi_{\mS^3}(t,r) = -\frac{GM}{a}\cot\left(r\sqrt{\CR/6}\right)\left[1-\left(r\sqrt{\CR/6}\right)/\pi\right],
\end{align}
setting $A = 0$. Then, in the limit $\CR \rightarrow 0^+$ we obtain $\Phi(r) = -\frac{GM}{a r}$ which is Newton's law with expansion.\footnote{In the limit, using Eq.~\eqref{eq::NEN_exp_2}, the expansion actually disappears since we obtain an infinite Universe with a finite mass.}

This limit is, of course, defined for $\CR > 0$, not at $\CR = 0$ itself, where the topology is changed. The limit is also defined, and the result equivalent, if we consider $r \rightarrow 0^+$ instead of $\CR \rightarrow 0^+$, as done with the Taylor series. In other words, the form of the Green function solution of the (cosmological) Poisson equation around the singularity does not depend on the boundary conditions, i.e. the topology, even though these conditions are needed to calculate that function.

The fact that in the small scale limit $r \rightarrow 0^+$, the potential is equivalent for any topology (as can be seen in the $\Phi_{-1}$ column of Table~\ref{tab:results}), implies that the modification of the large-scale properties of space only has weak effects on small-scale gravitational effects.

\section{Discussions}
\label{sec:disc}

One of the main purposes of the formulae derived in Secs.~\ref{sec:grav_sys_pm} and~\ref{sec:N-body} is to provide a framework to perform $N$-body simulations of structure formation in spherical topologies. With Newton's theory (i.e. in a Euclidean topology), such simulations are mainly used to study the non-linear regime of structure formation, while the linear regime is usually described analytically with the weak-field limit of general relativity. Since the scales of non-linearities ($\lesssim 10$ Mpc), as well as the largest linear inhomogeneity scale ($\sim 100$ Mpc), are small compared to typical estimates ($\gtrsim 10$ Gpc) of lower bounds (in certain cases) of the finite size of our Universe and its curvature radius, we expect the effects of topology and curvature on (non)-linear structures to be weak.

  We showed above in Table~\ref{tab:results} that the lowest order terms of the Taylor series of the potential are isotropic. Thus, an alternative numerical strategy to using the exact expressions of Secs.~\ref{sec:grav_sys_pm} and~\ref{sec:N-body} would be to use the lower orders of the Taylor series. This would be justified to third order.

  Independently of numerical strategies, these expansions help to understand the roles of curvature and topology, that are to some degree separated. This is discussed in the following sections.

\subsection{Isotropic terms}
\label{sec:iso}

Table~\ref{tab:results} shows the terms of the expansion series of each regular spherical topology through to the highest isotropic term, i.e. that does not depend on $\theta$ or $\varphi$. For $\mS^3$, i.e. the 1-cell topology, the solution is formula~\eqref{eq:phi_sol_S3} and is therefore isotropic at full order. This is not the case for the other regular spherical topologies, where the isotropic property of the gravitational potential is violated at a high order, depending on the topology. The Poincar\'e space, which tiles $\mS^3$ with 120 cells, is the most isotropic space, in the sense that the potential remains isotropic up to and including the fifth order, which corresponds to the fourth order for the gravitational field $\T g \coloneqq -\T D \Phi$.

This newly found uniqueness of the Poincar\'e space is qualitatively similar to that found with the earlier, adjacent-images heuristical approach, in which the Poincar\'e space was the ``best-balanced'' \cite{2009_Roukema_et_al}, but is better justified physically using the current approach. What also remains qualitatively confirmed in the study of topological acceleration \cite{2007_Roukema_et_al} is that the local kinematics and the integrated spacetime paths of extragalactic objects carry, in principle, information that characterises the global topology of the Universe.

We expect anisotropic terms for spherical topologies to be generically much more common than the isotropic terms.
However, averaging of observations under the assumption of intrinsic isotropy often enables the extraction of information with a minimum of free parameters: it will generally be easier to infer isotropic terms than anisotropic ones.
Nevertheless, investigating if these anisotropic terms are useful for distinguishing different topologies would be worth followup work.

\subsection{Interpretation of the even and odd orders}
\label{sec:disc_even_odd}

In Table~\ref{tab:results}, we see that the spherical topologies have the same isotropic odd orders of their expansion series at fixed curvature, and the same even orders at fixed volume. For the 3-torus, the first order is missing, in contrast to the spherical cases, but again, the second order is the same at fixed volume. To interpret this remarkable feature let us consider the solution of the Poisson Eq.~\eqref{eq::NEN_Poisson} in $\mH^3$ (we discuss the physical relevance of this solution in the Appendix), which can be thought of as $\mathbb{R}^3$ on which a non-zero spatial curvature of the form $\CR_{ij} = \left(\initial{\CR}/ (3a^2)\right)\, h_{ij} = -(2/a^2)\, h_{ij}$ is imposed: we have $D_cD^c\Phi_{\mH^3} = 4\pi GM\delta_{\mH^3}$, which leads to
\begin{align}
        \partial_\xi^2 \Phi_{\mH^3} + 2\coth\xi \, \partial_\xi \Phi_{\mH^3} = 4\pi GM\delta_{\mH^3},
\end{align}
where $\sinn\left(\xi\right) = \sinh\left(\xi\right)$ in formula~\eqref{eq::sinn}. The solution is
\begin{align}
  \Phi_{\mH^3} = - \frac{GM}{a(t)}\coth(\xi).
  \label{eq:Phi_hyperbolic}
\end{align}
We give the expansion series around $\xi\sim0$ of this solution, as a function of the distance $r = a\xi$ from the origin, in the last line of Table~\ref{tab:results}. The odd positive orders are the same as in the spherical topologies when normalised by the scalar curvature. However, the even terms are missing.

Thus, it appears that we can interpret the odd (positive) orders as an effect of non-zero spatial curvature, as these are present for the spherical topologies, but are absent for $\mathbb{T}^3$ and $\mE^3$, where $\CR = 0$.
Moreover, we can interpret the even orders to be an effect of the closedness (volume finiteness, in this context) of the manifold, as these are missing for $\mH^3$ and $\mE^3$ (which are both open manifolds), but are present for all the spherical topologies (which are necessarily closed) and for~$\mathbb{T}^3$. However, while the odd terms depend solely on the curvature, the even terms can depend on both, and not solely on the volume. This is the case for the fourth (isotropic) order.

In summary, it appears that, apart from the classical term in $1/r$, the odd (isotropic) orders indicate curvature and the even (isotropic) orders indicate finiteness.

\subsection{Dominant effects: Curvature versus topology}
\label{sec:dominant}

For a fixed curvature, Table~\ref{tab:results} shows that the gravitational potential starts differing among the spherical topologies at the second order ($V_\Sigma$ depends on the number of images, and thus on the holonomy group $\Gamma$). This implies that in cases where the first order term of the Taylor series dominates, the departure from the classical Newtonian law $1/r$ is mainly controlled by the spatial curvature, and the specific choice of topology has a subdominant effect on the gravitational potential\footnote{This result might not hold for globally inhomogeneous topologies.}. Since the inhomogeneity length scale of cosmic structures is small compared to the minimal size of the Universe typically inferred from WMAP and Planck data \citep{2014_Roukema_et_al,2015_Planck_XVIII}, the first few orders of the Taylor series will tend to be sufficient to describe curvature and topological effects on the gravitational potential and structure formation.

We estimate the correction to the classical Newtonian law that can be attributed to spatial curvature and topology by evaluating $\Phi_1/\Phi_{-1}$ and $\Phi_2/\Phi_{-1}$ for the homogeneity scale ($r_h \sim 100$~Mpc$/h$) and restoring the powers of $r$. We consider the value of the curvature parameter $|\Omega_k| \coloneqq |\CR|c^2/(6H_0^2) \sim 0.05$, currently given by studies of the CMB inferring a positive curvature (e.g. \citep{2020_Di-Valentino_et_al}), i.e. $|\CR| \approx 0.033 h^2$~Gpc$^{-2}$. This is not the usually accepted value, but is the highest estimate of non-negligible spatial curvature currently debated. Thus, for the four spherical topologies we have
\begin{align}
  \left|\frac{\Phi_1 \,r_h}{\Phi_{-1}/r_h}\right| & \approx 1.9 \times 10^{-5}  \quad {\rm for}\quad \mS^3, \, M_3, \, M_6, \, M_7,
 \end{align}
 and
 \begin{align}
   \left|\frac{\Phi_2 \,r_h^2}{\Phi_{-1}/r_h}\right| & \approx
  \begin{cases}
    4.4 \times 10^{-8} & \mS^3 \\
    3.5 \times 10^{-7} & M_3 \\
    1.1 \times 10^{-6} & M_6 \\
    5.3 \times 10^{-6} & M_7
  \end{cases}
  \,,
\end{align}
independently of $H_0$.
In the case of the Poincar\'e space, the second order term is only four times weaker than the first order, making the separation between curvature and topological effects more difficult than for the other topologies, in which the second order is even weaker. The third and fourth order terms are several orders of magnitude weaker.
Nevertheless, the overall amplitudes are weak compared to the classical Newtonian term. On typical scales of interest, the effects of spatial curvature or topology that are directly detectable from high accuracy estimates of the gravitational potential are likely to be very weak. Still curvature may be probed via large scale structure observations by several other methods, such as the clustering ratio recently proposed in Ref.~\citep{2022_Bel_et_al}.

What has a better prospect of detectability is that the long-term effect of $\Phi_1$ and $\Phi_2$, integrated over gigayear timescales. This makes $N$-body simulations in a spherical or hyperbolic universe a relevant study that might be able to constrain global curvature, topology or both. For such a study, the NEN theory in the form presented in Sec.~\ref{sec:N-body} provides an ideal mathematical tool.

\section{Conclusion}
\label{sec:ccl}

In this paper we used the non-Euclidean Newtonian theory developed in \cite{2022_Vigneron_b} to study non-relativistic effects of spherical topologies on the gravitational potential. We provided the general formula for the potential in any spherical topology [Eq.~\eqref{eq:Phi_app}]. We calculated its Taylor series near a point mass in the globally homogeneous ``regular'' spherical topologies (the geometrically simplest spherical topologies). The results are summarised in Table~\ref{tab:results}. Since the size of the cosmic structures are expected to be small compared to the curvature radius or the finite size of the Universe, the first orders of this Taylor series provide a good estimation of the gravitational effects that curvature and topology should have on structures.

As in the case of the (Euclidean) 3-torus, the potential in spherical topologies includes terms in the Taylor series beyond the classical $1/r$ term, for which we propose an interpretation of the different orders: (i) the isotropic even orders can be interpreted as an effect of the closedness of the manifold; (ii) the isotropic odd orders can be interpreted as an effect of non-zero spatial scalar curvature. A consequence is that, compared to the point mass solution in a 3-torus, widely used in Newtonian cosmological simulations, the spherical cases all feature an additional attractive first order term dependent solely on the spatial curvature. We also showed that the effect of the choice of topology is moderately weaker than that of global spatial curvature for the Poincar\'e dodecahedral space, $M_7$, and significantly weaker than the effect of curvature for the other three spherical spaces. This suggests that topology and curvature should have separable effects on the dynamics of structures.

We provided formulas to be used to perform $N$-body simulations aiming at studying structure formation in spherical topologies (Sec.~\ref{sec:N-body}). The main concern for performing these simulations is that the current constraints given by interpreting the Planck data and baryon acoustic oscillation measurements within the homogeneous and isotropic Friedmann models \citep{2020_Planck_VI} imply a negligible spatial curvature ($|\Omega_k| \lesssim 10^{-3}$). However, there has been a growing debate over the past few years whether or not the CMB power spectrum alone favours positive curvature (e.g. \cite{2020_Di-Valentino_et_al, 2021_Handley}). A structure formation simulation in a spherical universe under the assumption of the NEN theory would test global curvature both by its usual Friedmannian effects and by its effects in the first order of the potential: the two effects would have to agree on the value of the scalar curvature $\CR$. The NEN theory should thus lead to experimentally falsifiable predictions.

Finally, in this paper we only considered spherical topologies, and we focused on the spaces most likely to have isotropic effects. Calculating the gravitational potential in multiply connected hyperbolic 3-manifolds is left to future work. Generalising the non-Euclidean Newtonian theory to all of the topologies of the Thurston classification would also be an interesting study that would provide a more complete understanding of topological effects in cosmology.\saut

\textit{Data and code availability} -- The scripts for calculating and confirming the results in Tables~\ref{tab:results}, \ref{tab:results_A}, and \ref{tab:results_Phi_0} are available as free-licensed software (GPL-2 or later) at \url{https://codeberg.org/boud/topoaccel}.
These can be run using the free-licensed software package {\sc Maxima} (\mbox{\url{https://maxima.sourceforge.io/documentation.html}}).

\section*{Acknowledgements}
Part of this work has been supported by the Polish MNiSW Grant No. DIR/WK/2018/12.
Part of this work has been supported by the Pozna\'n Supercomputing and Networking Center (PSNC) computational Grant No. 537. Q.V. was supported by the Centre of Excellence in Astrophysics and Astrochemistry of Nicolaus Copernicus University in Toru\'n, and by the Polish National Science Centre under Grant No. 2022/44/C/ST9/00078. We thank Etienne Jaupart for useful discussions.

\appendix

\section{Is the solution in $\mH^3$ physical?}
\label{sec:disc_H3}

While of interest in the current work, the solution of the Poisson equation in the infinite space $\mH^3$ would not normally be considered to be physical as a non-Euclidean {Newtonian} gravitational potential  in the sense of NEN theory, which prioritises topological classification over geometrical properties. Thus, for two manifolds having the same topology, only one non-relativistic (in other words Newtonian-like) theory should be considered physically valid. In particular, if the topology of the manifold is that of $\mE^3$, we should necessarily take a zero Ricci tensor, and use (Euclidean) Newton theory. So, even though $\mH^3$ and $\mE^3$ are not the same Riemannian manifold in the sense that the Riemann structures defined on the manifolds are different, they are the same topological manifold. Therefore, there exists only one Newtonian gravitational field, which is that given by considering the Ricci tensor to be zero. Thus, the solution of the equation $D_cD^c\phi = 4\pi GM\rho$ with $\CR_{ij} = (\CR/3)\, h_{ij}$ and $\CR <0$ should not normally be considered as the gravitational potential in the corresponding topological space.

For clarification, the procedure for calculating the gravitational field in a 3-manifold in NEN theory is the following:
\begin{enumerate}
        \item We choose the topology of the manifold $\Sigma$ in which we want to calculate the gravitational field.
        \item Following the procedure proposed in \cite{2022_Vigneron_b}, the Ricci tensor that needs to be considered should be the ``simplest'' one that can be defined in the topological space $\Sigma$. If the topology is irreducible in the sense given by the Thurston decomposition (necessarily closed, i.e. of finite volume), then $\CR_{ij}$ is given by the spatial metrics in \cite{1995_La_Lu}; if instead the topology is that of $\mathbb{R}^3$, then one must take $\CR_{ij} =0$.
\end{enumerate}
Thus, this procedure excludes a Riemannian manifold which has the topology of $\mathbb{R}^3$ but a non-zero Ricci tensor, i.e. we cannot have $\mH^3$, and instead we only have $\mE^3$. This is a consequence of hyperbolic topologies in the Thurston classification only including closed manifolds. For closed hyperbolic 3-manifolds, the equation $D_cD^c\phi = 4\pi GM\left(\delta_\Sigma - M/V_\Sigma\right)$ is valid, with $\CR_{ij} = \CR/3 h_{ij}$ and $\CR < 0$, since $V_\Sigma$ is defined (finite).\saut

\remark{The splitting~\eqref{eq:splitting} is only possible because the volume of the covering space $\mS^3$ is finite. For Euclidean or hyperbolic topologies, performing this decomposition ``naively'' would lead to an infinite, divergent sum. In these two cases, a method to enable the calculation of the potential is to renormalise the divergent sum (see~\cite{2016_Steiner} for the case of a 3-torus and the Poisson equation using the absolute density). For example, the Ewald summation used in some $N$-body codes corresponds to such a renormalisation in the case of the (Euclidean) 3-torus.}

\section*{References}
 {\small When available this bibliography style features three different links associated with three different colors: links to the journal/editor website or to a numerical version of the paper are in \textcolor{LinkJournal}{red}, links to the ADS website are in \textcolor{LinkADS}{blue} and links to the arXiv website are in \textcolor{LinkArXiv}{green}.}\\

\IfFileExists{QV_mnras.bst}{\bibliographystyle{QV_mnras}}{\bibliographystyle{/Users/quentinvigneron/Documents/Travail/Research/tex_/QV_mnras}}
\IfFileExists{bib_General.bib}{\bibliography{bib_General}}{\bibliography{/Users/quentinvigneron/Documents/Travail/Research/tex_/bib_General}}

\end{document}